\documentclass[11pt,a4paper]{article} 
\pdfoutput=1
\usepackage{jheppub}

\usepackage{amsmath, amssymb}
\usepackage{mathpazo}
\usepackage{mathrsfs}
\usepackage{array,arydshln}

\usepackage{graphicx,epsfig}
\usepackage{epic}
\usepackage{color}
\usepackage{youngtab}
\usepackage{float}

\newcommand{\be}{\begin{equation}}
\newcommand{\ee}{\end{equation}}
\newcommand{\ba}{\begin{eqnarray}}
\newcommand{\ea}{\end{eqnarray}}

\newcommand{\mc}{\mathcal }

\newcommand{\mk}{\mathfrak}

\def\XXint#1#2#3{{\setbox0=\hbox{$#1{#2#3}{\int}$}
     \vcenter{\hbox{$#2#3$}}\kern-.5\wd0}}

    \newcommand{\beq}{\begin{equation}}
    \newcommand{\eeq}{\end{equation}}
    \newcommand\beqa{\begin{eqnarray}}
    \newcommand\eeqa{\end{eqnarray}}

\title{On the one-loop curvature function in the $\mk{sl}(2)$ sector of $\mc N=4$ SYM}

\author[a,b]{Matteo Beccaria} 
\author[a,b]{, Guido Macorini} 

\affiliation[a]{Dipartimento di Matematica e Fisica Ennio De Giorgi,\\
Universit\`a del Salento \& INFN, Via Arnesano, 73100 Lecce, 
Italy} 

\affiliation[b]{INFN, Sezione di Lecce, Via Arnesano, 73100 Lecce, 
Italy} 
                     
\emailAdd{matteo.beccaria@le.infn.it}
\emailAdd{guido.macorini@le.infn.it}

\abstract{
We consider twist $J$ operators with spin $S$ in the $\mk{sl}(2)$ sector of $\mc N=4$
SYM. The small spin expansion of their anomalous dimension defines the so-called slope
functions. Much is known about the linear term, but the study of the quadratic correction, the curvature function,
started only very recently. At any fixed $J$, the curvature function can be extracted  at all loops from the 
$\mathbf{P}\mu$-system formulation of the Thermodynamical Bethe Ansatz. 
Here, we work at the one-loop level and follow a different approach. We present 
a systematic double expansion of the Bethe Ansatz equations at large $J$ and small winding number. 
We succeed in fully resumming this expansion and obtain a closed explicit simple formula for the 
one-loop curvature function. The formula is parametric in $J$ and can be evaluated with 
minor effort for any fixed $J$. The result is an explicit series in odd-index $\zeta$ values.
Our approach provides a complete reconciliation 
between the $\mathbf{P}\mu$-system predictions and the large $J$ approach.
}

\allowdisplaybreaks

\begin{document} \maketitle

\bigskip

\section{Introduction and summary of results}

The holographic duality between planar four dimensional N = 4 supersymmetric Yang- Mills theory (SYM) and string theory on $AdS_{5} \times S^{5}$ has been investigated to a remarkable level of accuracy thanks to the role
played by integrability on both sides of the correspondence \cite{Beisert:2010jr}. In particular, integrability
methods can be applied to the calculation of the anomalous dimensions of  (planar) single trace operators
as well as to the energy of the dual string states. In the integrability approach, the two quantities are essentially the same
object depending on the  't Hooft coupling $\lambda$ whose effects are treated non-perturbatively.

\vskip 5pt
In the large volume limit, the spectrum is captured by 
a set of asymptotic Bethe Ansatz equations \cite{Beisert:2005fw}. 
Finite size corrections are nowadays under full 
control by means of the Thermodynamical Bethe Ansatz (TBA) machinery, an infinite set of integral equations 
\cite{Gromov:2009tv,Bombardelli:2009ns,Gromov:2009bc,Arutyunov:2009ur,Cavaglia:2010nm}.
Recently, the TBA equations have been recast in the so-called quantum spectral curve or $\mathbf{P}\mu$-system
\cite{Gromov:2013pga}. This new proposal is a nonlinear Riemann-Hilbert problem for a set of only a few functions
and is much simpler than the original formulation. A remarkable application of the $\mathbf{P}\mu$-system
can be found in \cite{Gromov:2013qga}. Extensions to the ABJM theory are discussed 
in \cite{Gromov:2014eha,Cavaglia:2014exa}.

\vskip 5pt
In this paper, we focus on the $\mathbf{P}\mu$-system prediction for the so-called curvature function
associated with twist $J$ operators with spin $S$ in the $\mk{sl}(2)$ sector of $\mc N=4$ SYM. 
Such operators have the generic form 
\be
\mc O_{J, S}^{I} = \mbox{Tr}\bigg(Z^{J-1}\,D^{S}\,Z\bigg)+\cdots,
\ee 
where $Z$ denotes one of the complex scalars of the theory, $D$ is a light-cone covariant derivative
and the dots stand for permutations required in order to build a dilatation operator eigenstate, labeled by $I$.
In the following, we shall omit this index because we shall consider the state with minimal scaling dimension $\Delta$.  
As usual, the scaling  dimension of $\mc O$ is split into a classical plus quantum part 
\be
\Delta = J+S+\gamma(S,J; g),\qquad g = \frac{\sqrt\lambda}{4\,\pi}.
\ee
The anomalous dimension $\gamma(S, J; g)$ admits a small $S$ expansion
\be
\label{eq:smallS}
\gamma(S,J;g) = \gamma^{(1)}(J; g)\,S+\gamma^{(2)}(J; g)\,S^{2}+\cdots.
\ee
The first term is called the {\em slope function}, while the second has been recently dubbed the
 {\em curvature functions} in \cite{Gromov:2014bva}. Expansions with respect to charges like $S$ or $J$ are
 quite interesting since their coefficients are functions of $\lambda$ that can be studied both at weak and strong
 coupling, {\em i.e.} in the gauge or string theory. Therefore, any integrability based calculation that is able to interpolate
 between small and large values of $\lambda$ becomes immediately a test of the holographic correspondence~\footnote{Another celebrated example is the cusp anomalous dimension, a.k.a. the scaling function,  appearing as the coefficient of the leading $\log S$ term at large $S$, see for instance \cite{Freyhult:2010kc}.}.
 
\vskip 5pt
Since $S$ is integer, the expansion (\ref{eq:smallS}) is mathematically ill defined and deserves some caution.
Nevertheless, based on various physical assumptions about the solutions of the Bethe equations, 
the slope function has been determined in closed form
at all loops \cite{Basso:2011rs,Basso:2012ex,Gromov:2012eg}, and  reads
\be
\label{eq:slope}
\gamma^{(1)}(J; g) = \frac{4\,\pi\,g}{J}\frac{I_{J+1}(4\,\pi\,g)}{I_{J}(4\,\pi\,g)} = \frac{8\,\pi^{2}\,g^{2}}{J(J+1)}
-\frac{32\,\pi^{4}\,g^{4}}{J(J+1)^{2}(J+2)}+\cdots ,
\ee
where $I_{J}$ are modified Bessel's functions. This quantity is protected from wrapping corrections and is also 
insensitive to the dressing phase in the asymptotic Bethe Ansatz \cite{Vieira:2010kb} that contributes at $\mc O(S^{2})$.

\vskip 5pt
Going to the next order in the small $S$ expansion, 
the curvature function has been computed recently in \cite{Gromov:2014bva} by means of the 
$\mathbf{P}\mu$-system at all loops. The curvature function is somewhat richer than the linear slope because 
it receives contributions from the dressing phase and from wrapping corrections. The weak-coupling expansion 
of the curvature function has been reported in  \cite{Gromov:2014bva}  for specific values of the 
twist $J=2, 3, 4$. In principle, it is possible to evaluate it at any integer $J$ although it appears that the calculation is 
more and more cumbersome as $J$ increases.

\vskip 5pt
A quite different approach to the analysis of (\ref{eq:smallS})  has been pursued in 
\cite{Beccaria:2012kp} in the context of the investigation
of the spectrum of semiclassical quantum strings in $AdS_{5}\times S^{5}$ on the example of folded $(S,J)$ 
string (with spin $S$ in $AdS_{5}$ and orbital momentum $J$ in $S^{5}$) dual to the above gauge theory states.
One of the outcomes of the analysis of  \cite{Beccaria:2012kp} is that the weak-coupling anomalous dimension
$\gamma(S,J; g)$ can be expanded at large $J$ in powers of $1/J$ with coefficients that are polynomials in $S$. 
Thus, it is possible to give a non ambiguous meaning to (\ref{eq:smallS}), at least order by order in $1/J$. 
Wrapping corrections are lost in this approach, since they are exponentially suppressed at large $J$. Nevertheless, 
this is not a problem for the linear slope, which is independent on wrapping, as well as for the higher order slopes
at one loop. 

\vskip 5pt
At the leading order in $S$, one easily checks that the large $J$ expansion agrees at one-loop 
with the first term in (\ref{eq:slope}) that is $\sim 1/J(J+1)$.
On the other hand, for the curvature function, an expansion like (\ref{eq:slope}) is not available. In particular, 
we don't control parametrically the dependence on $J$. We only have the weak-coupling expansion of the curvature
function at specific values of $J$ from the $\mathbf{P}\mu$-system \cite{Gromov:2014bva}.
Matching the large $J$ expansion is then non trivial, since one needs the resummation of an infinite 
series in $1/J$. 

\medskip
The problem is difficult  even at one-loop. In more details, we can split the one-loop anomalous dimension
and write
\be
\gamma^{(2)}(J; g) = g^{2}\,\gamma^{(2)}_{1}(J)+\mc O(g^{4}).
\ee
In \cite{Gromov:2014bva}, the values of $\gamma^{(2)}_{1}(J)$ are computed at $J=2,3,4$ and 
read~\footnote{Notice that the predictions at $J=2,3$ have been compared with the direct small $S$ expansion of 
the known analytical expressions of the one loop anomalous dimension. However, for $J\ge 4$, such a prediction is not available.}
\be
\label{eq:pmu-prediction}
\gamma^{(2)}_{1}(2) = -8\,\zeta_{3},\quad
\gamma^{(2)}_{1}(3) = -2\,\zeta_{3},\quad
\gamma^{(2)}_{1}(4) = -\frac{14}{5}\,\zeta_{3}+\frac{48}{\pi^{2}}\,\zeta_{5}-\frac{252}{\pi^{4}}\,\zeta_{7}.
\ee
These values should match the following infinite sum, taken from \cite{Beccaria:2012kp}, evaluated at the finite 
points $J=2,3,4$
\be
\label{eq:oldexp}
\gamma^{(2)}_{1}(J) = 16\,\pi^{2}\bigg(
-\frac{1}{4 J^3}+\frac{\frac{1}{8}-\frac{\pi ^2}{12}}{J^4}+\frac{-\frac{3}{16}+\frac{\pi
   ^2}{4}-\frac{\pi ^4}{90}}{J^5}+\frac{\frac{5}{32}-\frac{19 \pi ^2}{48}+\frac{2 \pi
   ^4}{45}-\frac{\pi ^6}{315}}{J^6}+\cdots
\bigg).
\ee
The problem with (\ref{eq:oldexp}) is that its terms have been computed by a semi-analytical method
and there is no control over their general structure. A few more terms can be added, as in \cite{Gromov:2014bva}. They lead to a surprisingly good numerical agreement for $J=4$, but work badly for $J=2,3$. 
All in all, the precise matching between explicit numbers like (\ref{eq:pmu-prediction}) and (\ref{eq:oldexp})
for generic $J$ remains until now an open problem.

\vskip 5pt
In this paper, we address the problem of reconciling (\ref{eq:oldexp}) with predictions like (\ref{eq:pmu-prediction})
in a general way. Our analysis will be based on a double expansion of the curvature function at 
large $J$ and small mode number, where the latter is a useful device to organise the various powers of $\pi$ in 
(\ref{eq:oldexp}). We shall provide strong arguments for the following compact formula providing the 
one loop curvature function at any $J$
\ba
\label{eq:intro-main}
\gamma_{1}^{(2)}(J) &=& \frac{8\pi^{3}}{J(J+1)}\int_{0}^{1}dx\,\bigg[x\,(1-x)^{2J-2}-\frac{x^{2}}{2}(1-x)^{J-1}\bigg(
(1+x)^{J-1}+(1-x)^{J-1}
\bigg)\bigg]\times \nonumber \\
&&\times \bigg(\tan\frac{\pi x}{2}-\cot\frac{\pi x}{2}\bigg).
\ea
This formula is the required bridge between the $\mathbf{P}\mu$-system and 
large $J$ approaches. Indeed, expanding the last factor inside the integral and performing the integration over $x$, 
one finds indeed
\ba
\label{eq:intro-analytic-resummed}
\gamma_{1}^{(2)}(J) &=& \frac{16\,\pi^{2}}{J(J+1)}\sum_{k=0}^{\infty}\bigg[
\frac{\Gamma(J)\,\Gamma(k+1)}{4\,\Gamma(J+k+1)}-\frac{(J-1)(4J+2k-1)\,\Gamma(2J-2)\,\Gamma(2k+1)}
{\Gamma(2J+2k+1)}
\bigg]\times \nonumber \\
&& \times (-1)^{k}\frac{2^{2k}}{(2k)!}B_{2k}\pi^{2k}.
\ea
Each term of this series is a rational function of $J$ and expanding at large $J$ one recovers (\ref{eq:oldexp}),
as we checked at order $1/J^{12}$. Also, evaluating (\ref{eq:intro-main}) at integer $J\ge 2$ one finds
the following finite sums of odd index $\zeta$ values
\ba
\label{eq:main-zeta}
\gamma_{1}^{(2)}(J) &=& \frac{16\,\pi^{3}}{J(J+1)}\sum_{n=0}^{J-1}\sum_{k=0}^{J}\sum_{u=3}^{k+n+1}
(-1)^{k+n+u}\sin\left(\frac{\pi\,u}{2}\right)\,2^{-u}\frac{u!}{J\,(k+n+2)}\times\nonumber \\
&&\times
(k-2\,J+k\,(-1)^{k})\,
\binom{J-1}{n}\,\binom{J}{k}\,\binom{k+n+2}{u}\,\frac{\zeta_{u}}{\pi^{u}}.
\ea
Setting $J=2,3,4$ , the results (\ref{eq:pmu-prediction}) are obtained (for additional explicit points, see Sec.~(\ref{sec:match})).

\vskip 5pt
In more details, the plan of the paper is the following. 
In Sec.~(\ref{sec:BA}), we formulate the relevant Bethe Ansatz problem and present various tools to derive analytically
its
expansion at large $J$ and small winding. 
In Sec.~(\ref{sec:resummation}), we resum the $J$ dependence of the curvature function, order by order in the small
winding parameter. 
In Sec.~(\ref{sec:match}), we further resum the dependence on the winding parameter, thus arriving at our proposed
closed formula.
App.~(\ref{app:explicit}) collects long explicit expansion that extend the data available in literature. 
App.~(\ref{app:convergence}) proves a convergence property of the small winding expansion in rigorous way.

\section{One loop Bethe Ansatz equations and their expansion}
\label{sec:BA}

In the following, $J$ is a positive number and $S$ is a positive even integer. The physically relevant case is 
$J$ integer $\ge 2$. About $S$, the results will be polynomial in $S$ and will be valid for any integer $S$. 
The choice $S$ even just simplifies the discussion.

The one-loop Bethe equations for twist $J$ operators with spin $S$ in the $\mk{sl}(2)$ sector
of $\mc N=4$ SYM are  
\be
\label{eq:Bethe}
-\varepsilon_{i}\,\pi+J\,\arctan\left(\frac{1}{2\,u_{i}}\right)+\sum_{j\neq i}^{S}
\arctan\left(\frac{1}{u_{i}-u_{j}}\right)=0,\qquad i=1, \dots, S,
\ee
where $\varepsilon_{i}=1$ for $i=1, \dots, \frac{S}{2}$, and $\varepsilon_{i}=-1$ for $i=\frac{S}{2}+1,\dots, S$.
The $\varepsilon_{i}$ term fixes the mode numbers of the solution to be those of the ground state of the associated spin-chain.  The solution is unique and symmetric under $u\to -u$. 
The  associated (one-loop) energy of the spin-chain is~\footnote{
The relation with $\gamma_{1}(S,J)$ in $\gamma(S,J;g) = g^{2}\,\gamma_{1}(S,J)+\cdots$
is simply $\gamma_{1}(S,J) = 16\,\pi^{2}\,E(S,J)$.
}
\be
E(S,J) =\frac{1}{8\,\pi^{2}}\, \sum_{i=1}^{S}\frac{1}{u_{i}^{2}+\frac{1}{4}} = 
\frac{1}{4\,\pi^{2}}\, \sum_{i=1}^{\frac{S}{2}}\frac{1}{u_{i}^{2}+\frac{1}{4}}.
\ee

At generic $S>4$, the energy is not a rational function of $J$. Nevertheless, the large $J$ expansion takes the following
form 
\be
E(S,J) = \sum_{n=2}^{\infty}\sum_{m=1}^{n-1}c_{n}^{(m)}\,\frac{S^{m}}{J^{n}} = E^{(1)}(J)\,S+
E^{(2)}(J)\,S^{2}+\dots, 
\ee
where the constants $c_{n}^{(m)}$ are rational combinations of even powers of $\pi$ up to $\pi^{2n-6}$. 
In App. (A), we report the extended $\mc O(1/J^{11})$ expansion
of $E^{(2)}(J)$ that we obtained by the semi-analitic method proposed in \cite{Beccaria:2012kp}. Inspection of the 
coefficients of the expansion shows that any simple structure is hindered by the presence of increasing powers of $\pi$.
These are not a mathematical artefact because $\pi$ enters the Bethe equations (only) in the first term of (\ref{eq:Bethe})
which is physically associated with the winding of the dual folded string. These qualitative remarks suggest a different 
way of organising the expansion of $E(S,J)$ that we present in the next section. This new expansion, in small
winding number, will turn out to be quite powerful as we shall see later.

\subsection{Small winding expansion}

Winding can be introduced by rescaling $\pi\to n\pi$ in the Bethe equations. Equivalently, one can 
reset $\pi\to P$ and consider an expansion in terms of the $P$ variable around $P=0$. In other words, we 
introduce the further expansion
\be
E^{(m)}(J) = \sum_{p=0}^{\infty}E^{(m)}_{p}(J)\,P^{2p}.
\ee
Let us begin with the leading order $E_{0}^{(m)}$. This is obtained by rescaling in the Bethe equations 
$u_{i}=x_{i}/P$ and taking $P\to 0$.
This leads to the simplified Bethe equations
\be
-\varepsilon_{i}+\frac{J}{2\,x_{i}}+\sum_{j\neq i}\frac{1}{x_{i}-x_{j}}=0, \qquad i=1, \dots, S.
\ee
The $\varepsilon_{i}$ term is physically very important and is the remnant of the fact that we are studying a 2-cut
solution of the Bethe equations in the continuum limit $S\to \infty$.  Compared to this case, the analysis 
of the similar equations for 1-cut solutions \cite{Lubcke:2004dg} is much simpler. The best way to treat the $\varepsilon$ term is to  
use the identity (a similar trick has been exploited in \cite{Beisert:2005mq})
\be
\int_{0}^{\infty} d\varepsilon \frac{x}{x^{2}+\varepsilon^{2}} = \frac{\pi}{2}\mbox{sign}(x).
\ee
Defining 
\be
G(x) = \sum_{i=1}^{S}\frac{1}{x-x_{i}}, \quad\mbox{and}\quad
\mc G(x) = J G(J x), 
\ee
we obtain after some manipulation~\footnote{We exploit in particular $G(x)=-G(-x)$ due to the symmetry $x_{i}\to -x_{i}$.}, for $x>0$~\footnote{For general $x$, the integral has a factor $\mbox{sign}(x)$ in front. In the following, we shall always take $x>0$.}
\be
\frac{1}{2J}(\mc G^{2}+\mc G')+\frac{\mc G(x)}{2\,x}
-\frac{2}{\pi}\int_{0}^{\infty} d\varepsilon \frac{\mc G(x)-i\,\varepsilon\,\mc G(i\,\varepsilon\,x)}{1+\varepsilon^{2}}=0.
\ee
Integrating the last term in the complex plane, this reduces to~\footnote{Here, there is an important hidden trick. In the large $J$ 
expansion, $\mc G(x)$ has poles only in $x=\pm 1/2$ order by order in $1/J$. Instead, at finite $J$, the function $\mc G(x)$ has poles at the scaled Bethe roots.}
\be
\frac{1}{2J}(\mc G'(x)+\mc G^{2}(x))+\frac{\mc G(x)}{2x}
-2\,i\,\mbox{Res}_{\varepsilon=\frac{i}{2x}}\frac{\mc G(x)-i\,\varepsilon\,\mc G(i\,\varepsilon\,x)}{1+\varepsilon^{2}}=0.
\ee
The residue can be further simplified and we arrive at the final form 
\be
\label{eq:master}
\frac{1}{2J}(\mc G'(x)+\mc G^{2}(x))+\frac{\mc G(x)}{2x}
-2\,\mbox{Res}_{u=1}\frac{u\,\mc G(u/2)}{4x^{2}-u^{2}}=0.
\ee
The energy is obtained from 
\be
\label{eq:energy}
E_{0}(J) = -\frac{1}{8J^{2}}\mc G'(0).
\ee
A detailed analysis of the large $J$ perturbative solution of this equation leads to the educated Ansatz
\be
\mc G(x) = \frac{4\,S\,x}{4x^{2}-1}+\sum_{n=1}^{\infty}\frac{1}{J^{n}}\,\frac{x}{(4x^{2}-1)^{2n+1}}
\sum_{k=0}^{2n-1} c_{n,k}x^{2k}.
\ee
Plugging this expansion in (\ref{eq:master}), we  determine systematically the coefficients $c_{n,k}$.
Many of them are are listed in App.~(\ref{app:LOc}).
Replacing these coefficients in the expression (\ref{eq:energy}) for $E_{0}(J)$, we obtain 
\ba
E_{0}(J) &=& \frac{S}{2 J^2}+\frac{-\frac{S^2}{4}-\frac{S}{2}}{J^3}+\frac{\frac{3
   S^3}{16}+\frac{S^2}{8}+\frac{S}{2}}{J^4}+\frac{-\frac{21 S^4}{128}+\frac{3
   S^3}{64}-\frac{3 S^2}{16}-\frac{S}{2}}{J^5}\nonumber \\
   &&+\frac{\frac{159
   S^5}{1024}-\frac{99 S^4}{512}+\frac{3 S^3}{16}+\frac{5
   S^2}{32}+\frac{S}{2}}{J^6}+\mc O\left(\frac{1}{J^{7}}\right).
\ea
in agreement with the $P\equiv\pi\to 0$ terms in App.~(\ref{app:explicit}).

\subsection{Next-to-leading order}

At the next-to-leading order $\mc O(P^{2})$ we face some technical difficulties that 
can be treated following the ideas in \cite{Astolfi:2008yw},
developed for the much simpler case of a one-cut solution.
To the next order in the small $P$ expansion we have 
\be
\label{eq:master}
0=-\frac{2}{\pi}\int_{0}^{\infty}d\varepsilon\frac{x_{i}}{x_{i}^{2}+\varepsilon^{2}}+\frac{J}{2\,x_{i}}+\sum_{j\neq i}\frac{1}{x_{i}-x_{j}}
+P^{2}\left(-\frac{J}{24\,x_{i}^{3}}-\frac{1}{3}\sum_{j\neq i}\frac{1}{(x_{i}-x_{j})^{3}}
\right)+\mc O(P^{3}).
\ee
After some technical manipulation collected in App.~(\ref{app:NLO}), this equation can be written 
\ba
\label{eq:master-NLO}
&& 0=\frac{1}{2J}(\mc G'(x)+\mc G^{2}(x))+\frac{\mc G(x)}{2x}
-2\,\mbox{Res}_{u=1}\frac{u\,\mc G(u/2)}{4x^{2}-u^{2}}+P^{2}\bigg[ \\
&&
-\frac{1}{24\,J^{2}}\frac{\mc G(x)-x\,\mc G'(0)}{x^{3}}-\frac{1}{3\,J^{3}}\,
\bigg(-\frac{1}{2}\mc H'(x)-\frac{1}{24}G'''(x)-\frac{1}{4}G'(x)^{2}\bigg)
\bigg]+\mc O(P^{3}),	\nonumber
\ea
where the function $\mc H(x)$ can be expressed in terms of $\mc G(x)$ as 
\ba
\mc H(x) &=& J\,\bigg(
2\,x\,\mbox{Res}_{u=1}\frac{\mc G(u/2)^{2}+\mc G'(u/2)}{4x^{2}-u^{2}}
-\frac{\mc G(x)^{2}+\mc G'(x)-\mc G'(0)}{4x}
\bigg)\nonumber\\
&&-\frac{1}{3}\,\mc G(x)^{3}-\mc G(x)\mc G'(x)-\frac{1}{3}\mc G''(x)+\mc O(P^{2}).
\ea
The energy has now an explicit $\mc O(P^{2})$ correction
\be
E_{0}(J)+E_{1}(J)\,P^{2} = -\frac{1}{8J^{2}}\bigg(\mc G'(0)-\frac{P^{2}}{24J^{2}}\mc G'''(0)\bigg).
\ee
The Ansatz for $\mc G(x)$ is the same as before, but now the coefficients $c_{n,k}$ have an additional 
term $\sim P^{2}$,
\be
c_{n,k}=c_{n,k; 0}+c_{n,k; 1}\,P^{2}.
\ee
The values $c_{n,k,0}$ are the previous ones. The new coefficients $c_{n,k,1}$ can be iteratively determined, as before, 
and are listed in App.~(\ref{app:NLOc}). Replacing in the expression for the energy, we find
\ba
&& E_{0}(J)+E_{1}(J)\,P^{2} =\frac{S}{2 J^2}+\frac{-\frac{S^2}{4}-\frac{S}{2}}{J^3}+\frac{\left(\frac{1}{8}-\frac{P^2}{12}\right)
   S^2+\frac{3 S^3}{16}+\frac{S}{2}}{J^4}\nonumber \\
   &&+\frac{\left(\frac{P^2}{24}+\frac{3}{64}\right)
   S^3+\left(\frac{P^2}{4}-\frac{3}{16}\right) S^2-\frac{21
   S^4}{128}-\frac{S}{2}}{J^5}+\frac{1}{J^{6}}\bigg(\left(\frac{1}{2}-\frac{5 P^4}{3}\right)
   S+\left(\frac{P^2}{384}-\frac{99}{512}\right) S^4\nonumber \\
   &&+\left(\frac{3}{16}-\frac{43 P^2}{192}\right)
   S^3+\left(\frac{P^4}{2}-\frac{19 P^2}{48}+\frac{5}{32}\right) S^2+\frac{159
   S^5}{1024}\bigg)+\frac{1}{J^{7}}\bigg(\left(25 P^4-\frac{1}{2}\right) S\nonumber \\
   &&+\left(\frac{1431}{4096}-\frac{13
   P^2}{256}\right) S^5+\left(\frac{25 P^2}{96}-\frac{765}{2048}\right)
   S^4+\left(\frac{P^4}{4}+\frac{15 P^2}{64}+\frac{21}{512}\right) S^3\nonumber \\
   &&+\left(-11 P^4+\frac{9
   P^2}{16}-\frac{11}{64}\right) S^2-\frac{315
   S^6}{2048}\bigg)+\mc O(1/J^{8}).
\ea
in agreement with the $P^{2}\equiv\pi^{2}$ terms in App.~(\ref{app:explicit}).

\section{Resummation with respect to $J$}
\label{sec:resummation}

The procedure outlined in the previous section can be  automatised. Doing so, we can extract, in particular, a long $1/J$ series for the $P=0$
coefficient of the quadratic slope $E_{0}^{(2)}(J)$. This can be resummed in the nice form 
\be
E_{0}^{(2)}(J) = -\frac{1}{2J(J+1)(2J-1)}.
\ee
The same approach can be applied to the next orders in the small $P$ expansion. Remarkably, in all the cases we worked out, it has been possible to resum the $1/J$ series. The final result turns out to be 
\ba
\label{eq:resummed}
&& E^{(2)}(J) = -\frac{1}{2 J (J+1) (2 J-1)}-\frac{2 J^2-2 J-1}{6 J^2 (J+1)^2 (2 J-1) (2 J+1)}\,\pi^{2}\nonumber \\
&& -\frac{4 J^2-2
   J-3}{45 J^2 (J+1)^2 (2 J-1) (2 J+1) (2 J+3)}\,\pi^{4}\nonumber \\
   &&-\frac{2 \left(8 J^4+32 J^3+28 J^2-38 J-45\right) }{315 J^2 (J+1)^2 (J+2) (J+3) (2 J-1) (2 J+1) (2 J+3) (2 J+5)}\,\pi^{6}\nonumber \\
   &&-\frac{4 \left(16 J^4+56 J^3+56 J^2-44 J-105\right)
   }{1575 J^2 (J+1)^2 (J+2) (J+3) (2 J-1) (2 J+1) (2 J+3) (2 J+5) (2 J+7)}\,\pi^{8} \\
   &&-\frac{8 \left(32 J^6+384 J^5+1640 J^4+2880 J^3+1478 J^2-2634 J-4725\right)}{6237 J^2 (J+1)^2 (J+2)
   (J+3) (J+4) (J+5) (2 J-1) (2 J+1) (2 J+3) (2 J+5) (2 J+7) (2 J+9)}\,\pi^{10}\nonumber\\
   &&+\dots\,,\nonumber
\ea
where dots stands for similar expressions, {\em i.e.} higher powers of $\pi$ times 
rational functions of $J$ that we did not compute (but see more later, in particular (\ref{eq:analytic-resummed})).

\vskip 5pt
We remark that we also tried to resum the coefficients of $P^{2k}$ in $E^{(3)}$, {\em i.e.} the cubic slope. Unfortunately, 
in this case, we could not find a simple resummation neither for the simplest coefficient which is that at $P=0$.
A similar negative result holds for the quadratic slope at two-loops. After some straightforward analysis, 
we found the following extension of the results of \cite{Beccaria:2012kp}  (see Eq.~(C-9) of that reference)
\ba
\left. \gamma_{2}^{(2)}(J)\right|_{\pi\to 0} &=& (16\,\pi^{2})^{2}\,\bigg(
-\frac{1}{4\,J^{5}}+\frac{5}{8\,J^{6}}-\frac{17}{16\,J^{7}}
+\frac{25}{16\,J^{8}}-\frac{131}{64\,J^{9}}+\frac{153}{64\,J^{10}}-\frac{595}{256\,J^{11}}
+\frac{337}{256\,J^{12}}\nonumber \\
&&+\frac{1741}{1024\,J^{13}} -\frac{9071}{1024\,J^{14}}+
\frac{100045}{4096\,J^{15}}+\cdots\bigg),
\ea
and, again, we could not resum this partial series by any simple rational Ansatz.

\subsection{Remarkable properties of the resummed small winding expansion}

Our result for $E_{0}^{(2)}(J)$ in (\ref{eq:resummed}) can be written in the much more suggestive form
that is obtained by factoring the leading term and replacing powers of $\pi$ by even argument $\zeta$ values
\ba
E_{0}^{(2)}(J) &=& -\frac{1}{2J(J+1)(2J-1)}\,\bigg[1+\nonumber\\
&&\bigg(
\frac{6}{J+1}-\frac{4}{2 J+1}-\frac{2}{J}
\bigg)\,\zeta_{2}+\nonumber \\
&&\bigg(
\frac{12}{J+1}+\frac{8}{2 J+1}-\frac{24}{2 J+3}-\frac{4}{J}
\bigg)\,\zeta_{4}+\nonumber \\
&&\bigg(
-\frac{6}{J+1}+\frac{30}{J+2}+\frac{14}{J+3}+\frac{36}{2 J+1}-\frac{40}{2
   J+3}-\frac{60}{2 J+5}-\frac{6}{J}
\bigg)\,\zeta_{6}+\nonumber \\
&&\bigg(
-\frac{72}{J+1}+\frac{40}{J+2}+\frac{168}{J+3}+\frac{80}{2
   J+1}+\frac{112}{2 J+3}-\frac{336}{2 J+5}-\frac{112}{2 J+7}-\frac{8}{J}
\bigg)\,\zeta_{8}+\nonumber \\
&&\bigg(
-\frac{210}{J+1}-\frac{260}{J+2}+\frac{700}{J+3}+\frac{270}{J+4}+\frac{22
   }{J+5}+\frac{140}{2 J+1}+\frac{720}{2 J+3}-\frac{504}{2
   J+5}\nonumber \\
   &&-\frac{1200}{2 J+7}-\frac{180}{2 J+9}-\frac{10}{J}
\bigg)\,\zeta_{10}+\cdots
\bigg].
\ea
This expression displays various intriguing regularities. First, all partial fractions have integer coefficient.
Then, we can identify general simple formulae for various terms inside the square bracket. Examples are the terms
\be
-\frac{2n}{J}\,\zeta_{2n},\quad
-\frac{2n(2n^{2}-6n+1)}{J+1}\,\zeta_{2n},\quad
\frac{4n(2n-3)}{2J+1}\,\zeta_{2n},\quad
\frac{4}{3}\frac{n(n-1)(2n-1)(2n-7)}{2J+3}\,\zeta_{2n}.
\ee
Finally, at large $J$, the leading term with coefficient $\zeta_{2n}$ is simply $2n/J^{n}$. All these remarkable
features suggest that a deeper understanding of the resummation (\ref{eq:resummed}) could be possible.
From a different perspectives, they support its validity. Indeed, we have to keep in mind that any resummation 
stands always as a conjecture because it is based on a finite number of terms, lacking a stronger analytical control
over the series that are resummed.

\section{Matching the one-loop $\mathbf{P}\mu$-system prediction at integer $J$}
\label{sec:match}

Let us begin with $J=2$. In order to match the $\mathbf{P}\mu$-system prediction (\ref{eq:pmu-prediction}),
we would like to show that 
\be
\label{eq:check}
\gamma_{1}^{(2)}(2) = -8\,\zeta_{3}.
\ee
Our resummed (with respect to $J$) expansion (\ref{eq:resummed}) reads instead
\be
\label{eq:J2-expansion}
\gamma_{1}^{(2)}(2) = -\frac{4 \pi ^2}{9}-\frac{2 \pi ^4}{135}-\frac{4 \pi ^6}{4725}-\frac{2 \pi
   ^8}{35721}-\frac{4 \pi ^{10}}{1002375}-\frac{4 \pi ^{12}}{13378365}+\cdots.
\ee
In order to understand how a series in $\pi$ can give a sum proportional to $\zeta_3$, it is natural to look for 
integral representations of $\zeta_{n}$ values containing $\pi$ as an explicit parameter. In particular, we remind that the following identities hold
\ba
\zeta(2n+1) &=& \frac{(-1)^{n}\,2^{2n}\,\pi^{2n+1}}{(2n+1)!}\int_{0}^{1}B_{2n+1}(x)\tan\frac{\pi x}{2}dx \nonumber \\
&=& 
 \frac{(-1)^{n-1}\,2^{2n}\,\pi^{2n+1}}{(2n+1)!}\int_{0}^{1}B_{2n+1}(x)\cot\frac{\pi x}{2}dx,
\ea
where $B_{n}(x)$ are Bernoulli polynomials.
We tried a simple linear combination of these identities and discovered that the series
(\ref{eq:J2-expansion}) is reproduced by the small $P$ expansion of the very simple combination
\be
F_{2}(P)=\frac{8P^{3}}{3}\int_{0}^{1}B_{3}(x)\bigg(\tan\frac{P x}{2}-\cot\frac{P x}{2}\bigg)\,dx,
\ee
where we remind that $B_{3}(x)=\frac{1}{2} (x-1) x (2 x-1)$ . 
Also, setting $P\to \pi$, we recover the $\mathbf{P}\mu$-system  result  (\ref{eq:check}) ! 
As an interesting  byproduct, 
the integral representation $F_{2}(P)$ allows to prove that the small winding expansion (\ref{eq:J2-expansion})
is convergent, see App.~(\ref{app:convergence}).

\vskip 5pt
The same analysis for $J=3$  leads to the combination 
\be
F_{3}(P)=-\frac{2P^{3}}{3}\int_{0}^{1}(1-x)^{2}\,x\,(x^{3}-x^{2}+3x-1)\bigg(\tan\frac{P x}{2}-\cot\frac{P x}{2}\bigg)\,dx,
\ee
Indeed, the small $P$ expansion of this quantity and its value at $P=\pi$ are
\ba
F_{3}(P) &=& -\frac{2 P^2}{15}-\frac{11 P ^4}{1890}-\frac{P ^6}{4725}-\frac{107 P
   ^8}{9823275}-\frac{82 P ^{10}}{127702575}-\frac{347 P ^{12}}{8428369950}+\cdots , \\
F_{3}(\pi) &=& -2\,\zeta_{3},
\ea
in agreement with both (\ref{eq:resummed}) and (\ref{eq:pmu-prediction}). Finally, for $J=4$, we found
the expression
\be
F_{4}(P) = -\frac{2P^{3}}{5}\int_{0}^{1}(1-x)^{3}\,x\,(4x^{3}-3x^{2}+4x-1)\bigg(\tan\frac{P x}{2}-\cot\frac{P x}{2}\bigg)\,dx,
\ee
and the data
\ba
F_{4}(P) &=& -\frac{2 P^2}{35}-\frac{23 P^4}{9450}-\frac{53 P^6}{779625}-\frac{23
   P^8}{7882875}-\frac{158 P^{10}}{1064188125}-\frac{401 P^{12}}{47760763050}+\cdots, \\
F_{4}(\pi) &=& -\frac{14}{5}\zeta_{3}+\frac{48}{\pi^{2}}\zeta_{5}-\frac{252}{\pi^{4}}\zeta_{7},
\ea
still in agreement with  (\ref{eq:resummed}) and (\ref{eq:pmu-prediction}).
A natural conjecture is then that the procedure can be extended to all $J$, with a function $F_{J}(P)$ 
of the form 
\be
\label{eq:result}
F_{J}(P) = P^{3}\int_{0}^{1}(1-x)^{J-1}\,x\,\mc F_{J}(x)\bigg(\tan\frac{P x}{2}-\cot\frac{P x}{2}\bigg)\,dx,
\ee
where $\mc F_{J}(x)$ is a polynomial with rational coefficients and degree $J$ for odd $J$, and $J-1$ for even $J$.
Our expansion (\ref{eq:resummed}) is too short to fix the coefficients of  $\mc F_{J}(x)$ for any $J$. Nevertheless,
we looked for solutions up to $J=8$ and found in all cases a unique solution with minimal coefficients, {\em i.e.}
without unnatural ugly rational coefficients. Besides, the found solution is quite regular and reads 
\ba
\mc F_{2}(x) &=& -\frac{4}{3}(2x-1), \\
\mc F_{3}(x) &=& -\frac{2}{3}(x^{3}-x^{2}+3x-1), \\
\mc F_{4}(x) &=& -\frac{2}{5}(4x^{3}-3x^{2}+4x-1), \\
\mc F_{5}(x) &=& -\frac{4}{15}(x^{5}-x^{4}+10x^{3}-6x^{2}+5x-1), \\
\mc F_{6}(x) &=& -\frac{4}{21}(6x^{5}-5x^{4}+20x^{3}-10x^{2}+6x-1), \\
\mc F_{7}(x) &=& -\frac{1}{7}(x^{7}-x^{6}+21x^{5}-15x^{4}+35x^{3}-15x^{2}+7x-1), \\
\mc F_{8}(x) &=& -\frac{1}{9}(8x^{7}-7x^{6}+56x^{5}-35x^{4}+56x^{3}-21x^{2}+8x-1). 
\ea
The strong regularity of the coefficients of these polynomials is remarkable. Indeed, 
one finds that in all cases 
\be
\mc F_{J}(x) = \frac{8}{J(J+1)}\sum_{k=0}^{k_{\max}}(-1)^{k}\,\frac{x^{k}}{k!}\left\{
\begin{array}{ll}
\frac{J!}{(J-k)!}, & \mbox{odd} \ J \\
\frac{(J-1)!}{(J-k-1)!}, & \mbox{even} \ J \\
\end{array}
\right.
\ee
where $k_{max} = J$ for odd $J$, and $J-1$ for even $J$. The sum can be done explicitly and we obtain 
our main result
\ba
\label{eq:main}
F_{J}(P) &=&  \frac{8P^{3}}{J(J+1)}\int_{0}^{1}dx\,\bigg[x\,(1-x)^{2J-2}-\frac{x^{2}}{2}(1-x)^{J-1}\bigg(
(1+x)^{J-1}+(1-x)^{J-1}
\bigg)\bigg]\times \nonumber \\
&&\times \bigg(\tan\frac{P x}{2}-\cot\frac{P x}{2}\bigg).
\ea
Clearing the smoke of our empirical derivation, one indeed checks that the expansion (\ref{eq:resummed})
is reproduced for any $J$ by expanding (\ref{eq:main}) in $P$ and integrating term by term. Together with the agreement with  (\ref{eq:pmu-prediction}) at $J=2,3,4$, this leads us 
to propose (\ref{eq:main}) at $P=\pi$ as the correct expression for the one-loop curvature function.
Evaluating it at various $J$, we obtain (\ref{eq:main-zeta}) in the Introduction.
Notice that the  integral in (\ref{eq:main}) can be quite efficiently determined by 
expressing, at each $J$,  the 
polynomial inside the integral as a linear combination of Bernoulli polynomials $B_{k}(x)$. Then, only 
odd index polynomials contribute through
\be
\int_{0}^{1}B_{k}(x) \bigg(\tan\frac{\pi x}{2}-\cot\frac{\pi x}{2}\bigg)dx = \frac{\zeta_{k}}{\pi^{k}}(-1)^{\frac{k-1}{2}}2^{2-k}k!\,,\qquad \mbox{odd}\ k,
\ee
while even index polynomials do not contribute due to $B_{2n}(x)=B_{2n}(1-x)$.
Specific cases of (\ref{eq:main-zeta}) are  
collected in the following list that extends  (\ref{eq:pmu-prediction})
\ba
\gamma_{1}^{(2)}(2) &=& -8 \zeta _3, \\
\gamma_{1}^{(2)}(3) &=& -2 \zeta _3\, \\
\gamma_{1}^{(2)}(4) &=& -\frac{14 \zeta _3}{5}+\frac{48 \zeta _5}{\pi
   ^2}-\frac{252 \zeta _7}{\pi ^4}, \\
\gamma_{1}^{(2)}(5) &=& -\frac{12 \zeta _3}{5}+\frac{48 \zeta _5}{\pi
   ^2}-\frac{252 \zeta _7}{\pi ^4}, \\
\gamma_{1}^{(2)}(6) &=& -\frac{44 \zeta _3}{21}+\frac{80 \zeta _5}{\pi
   ^2}-\frac{1860 \zeta _7}{\pi ^4}+\frac{21600 \zeta _9}{\pi ^6}-\frac{89100
   \zeta _{11}}{\pi ^8}, \\
\gamma_{1}^{(2)}(7) &=& -\frac{13 \zeta _3}{7}+\frac{750 \zeta _5}{7 \pi
   ^2}-\frac{6525 \zeta _7}{2 \pi ^4}+\frac{40500 \zeta _9}{\pi ^6}-\frac{334125
   \zeta _{11}}{2 \pi ^8}, \\
\gamma_{1}^{(2)}(8) &=& -\frac{5 \zeta _3}{3}+\frac{406 \zeta _5}{3 \pi
   ^2}-\frac{12565 \zeta _7}{2 \pi ^4}+\frac{184380 \zeta _9}{\pi
   ^6}-\frac{7397775 \zeta _{11}}{2 \pi ^8}\nonumber \\
   && +\frac{37837800 \zeta _{13}}{\pi
   ^{10}}-\frac{141891750 \zeta _{15}}{\pi ^{12}}.
\ea
%
It would be very interesting to check these results by the $\mathbf{P}\mu$-system at one-loop.

\subsection{A final non trivial consistency check}

Our main result (\ref{eq:main}) has been derived by some clever inspection of the data encoded in 
the expansion (\ref{eq:resummed}) where we have been able to determine the coefficients of the powers of $\pi$ up to 
$\pi^{10}$. Actually, the results in App.~(\ref{app:explicit}) contain various additional terms contributing the curvature
function with higher powers of $\pi$. A definitely non trivial consistency check amounts to re-derive the coefficients in 
(\ref{eq:resummed}) from (\ref{eq:main}) for higher powers of $\pi$ and compare their large $J$ expansion with 
App.~(\ref{app:explicit}). To this aim, we start from 
\be
\tan\frac{P x}{2}-\cot\frac{P x}{2} = \sum_{k=0}^{\infty} (-1)^{k-1}\frac{2^{4k}}{(2k)!}\,B_{2k}\,\bigg(\frac{P x}{2}\bigg)^{2k-1},
\ee
where $B_{2k}$ are Bernoulli numbers.
The term $\sim x^{k}$ can be plugged inside (\ref{eq:main}) replacing the tangent minus cotangent combination.
After integrating over $x$, this gives the coefficient of $\pi^{k+1}$ in the addition terms of (\ref{eq:resummed}).
The integration over $x$ can be done analytically for generic $J$ and we obtain 
the  following
closed expression for the series (\ref{eq:resummed}):
\ba
\label{eq:analytic-resummed}
E_{2}(J) &=& \frac{1}{J(J+1)}\sum_{k=0}^{\infty}\bigg[
\frac{\Gamma(J)\,\Gamma(k+1)}{4\,\Gamma(J+k+1)}-\frac{(J-1)(4J+2k-1)\,\Gamma(2J-2)\,\Gamma(2k+1)}
{\Gamma(2J+2k+1)}
\bigg]\times \nonumber \\
&& \times (-1)^{k}\frac{2^{2k}}{(2k)!}B_{2k}\pi^{2k},
\ea
which is (\ref{eq:intro-analytic-resummed}) mentioned in the Introduction.
In particular, we find the following additional terms to be added to those written in (\ref{eq:resummed})
\be
\frac{N_{12}(J)}{D_{12}(J)}\pi^{12}+\frac{N_{14}(J)}{D_{14}(J)}\pi^{14}+\frac{N_{16}(J)}{D_{16}(J)}\pi^{16}
+\cdots
\ee
where the explicit forms of the rational functions $N_{k}(J)/D_{k}(J)$ are
\ba
N_{12}(J) &=& -\frac{11056}{14189175}\,(64 J^6+736 J^5+3088 J^4+5272 J^3+3004 J^2-3254 J-10395), \nonumber\\
D_{12}(J) &=& J(J+1)\prod_{k=0}^{5}(J+k)\prod_{k=0}^{6}(2J+2k-1), \\
N_{14}(J) &=& -\frac{32}{57915}\,(128 J^8+3072 J^7+29568 J^6+145152 J^5+378672 J^4+479808 J^3\nonumber \\
&&+173752
   J^2-399342 J-945945), \nonumber \\
D_{14}(J) &=&  J(J+1)\prod_{k=0}^{7}(J+k)\prod_{k=0}^{7}(2J+2k-1), \\
N_{16}(J) &=& -\frac{231488}{516891375}\,(256 J^8+6016 J^7+57088 J^6+277120 J^5+717664 J^4+898384
   J^3\nonumber \\
   &&+357552 J^2-512280 J-2027025), \nonumber \\
D_{16}(J) &=& J(J+1)\prod_{k=0}^{7}(J+k)\prod_{k=0}^{8}(2J+2k-1).
\ea
The large $J$ expansion is thus
\ba
\frac{N_{12}(J)}{D_{12}(J)} &=& -\frac{5528}{14189175 J^9}+\frac{11056}{1289925 J^{10}}-\frac{176896}{1576575
   J^{11}}+\cdots, \\
   \frac{N_{14}(J)}{D_{14}(J)} &=& -\frac{16}{57915 J^{10}}+\frac{464}{57915 J^{11}}+\cdots, \\
   \frac{N_{16}(J)}{D_{16}(J)} &=& -\frac{115744}{516891375 J^{11}}+\cdots.
\ea
It can be compared with App.~(\ref{app:explicit}), upon identifying $P\equiv \pi$,  and perfect matching is found.

\section{Conclusions}

In this paper, we have presented a closed formula for the one-loop curvature function of 
twist $J$ operators in the $\mk{sl}(2)$ sector of $\mc N=4$ SYM. The formula agrees with the known 
prediction from the $\mathbf{P}\mu$-system as well as with the large $J$ expansion of the Bethe equations
that we have derived systematically. In this sense, it is a reconciliation between various approaches present in the 
literature about twist operators. The (one-loop) curvature function can be evaluated for any $J$ with minor effort. 
Various natural developments of the results presented in this paper are possible. In particular,
\begin{enumerate}
\item It would be interesting to derive  our main result (\ref{eq:intro-main}) by a weak-coupling expansion of the 
$\mathbf{P}\mu$-system. Indeed, the all-loop expressions of the curvature function from the $\mathbf{P}\mu$-system
have increasing complexity with $J$ and are not parametrical in this parameter. It would be nice to see how 
the simple dependence on $J$ that we presented can be extracted from the $\mathbf{P}\mu$-system.

\item It is intriguing that both the cubic slope and the two loop curvature functions are not covered by our analysis since their resummation, even at $P=0$, are not found in the class of rational functions of $J$. It could be that this fact is related 
to the role of the dressing phase. 

\item
Finally, the same methods could be applied to 
the ABJM theory where predictions from the $\mathbf{P}\mu$-system can also be extracted.
\end{enumerate}

\section*{Acknowledgments}

We thank G. Metafune for interesting discussions on the manuscript.

\appendix 

\section{Explicit expansions}
\label{app:explicit}

Let us define the polynomials
\be
P_{n}(S) = \sum_{m=1}^{n-1}c_{n}^{(m)}\,S^{m}.
\ee
Their explicit values for $n=2, \dots, 11$ are
\ba
P_2&=& \frac{S}{2},\\
P_3&=& -\frac{S^2}{4}-\frac{S}{2},\\
P_4&=& \frac{3
   S^3}{16}+(\frac{1}{8}-\frac{\pi ^2}{12}) S^2+\frac{S}{2},\\
P_5&=& -\frac{21
   S^4}{128}+(\frac{3}{64}+\frac{\pi ^2}{24}+\frac{\pi ^4}{180})
   S^3+(-\frac{3}{16}+\frac{\pi ^2}{4}-\frac{\pi ^4}{90}) S^2-\frac{S}{2},\\
P_6&=& \frac{159
   S^5}{1024}+(-\frac{99}{512}+\frac{\pi ^2}{384}-\frac{\pi ^4}{240}-\frac{\pi ^6}{1512})
   S^4+(\frac{3}{16}-\frac{43 \pi ^2}{192}-\frac{\pi ^4}{120}+\frac{11 \pi ^6}{3780})
   S^3\nonumber \\
   &&+(\frac{5}{32}-\frac{19 \pi ^2}{48}+\frac{2 \pi ^4}{45}-\frac{\pi ^6}{315})
   S^2+\frac{S}{2},\\
P_7&=& -\frac{315 S^6}{2048}+(\frac{1431}{4096}-\frac{13 \pi ^2}{256}+\frac{\pi
   ^4}{1920}+\frac{\pi ^6}{1512}+\frac{\pi ^8}{10800}) S^5+(-\frac{765}{2048}+\frac{25 \pi
   ^2}{96}+\frac{11 \pi ^4}{576}\nonumber \\
   &&+\frac{11 \pi ^6}{7560}-\frac{79 \pi ^8}{113400})
   S^4+(\frac{21}{512}+\frac{15 \pi ^2}{64}-\frac{\pi ^4}{48}-\frac{\pi ^6}{60}+\frac{47 \pi
   ^8}{28350}) S^3+(-\frac{11}{64}+\frac{9 \pi ^2}{16}-\frac{19 \pi ^4}{180}\nonumber \\
   &&+\frac{\pi
   ^6}{45}-\frac{2 \pi ^8}{1575}) S^2-\frac{S}{2},\\
P_8&=& \frac{321
   S^7}{2048}+(-\frac{17349}{32768}+\frac{41 \pi ^2}{384}+\frac{413 \pi ^4}{92160}-\frac{\pi
   ^6}{5376}-\frac{\pi ^8}{8640}-\frac{\pi ^{10}}{71280}) S^6+(\frac{12849}{16384}-\frac{1635
   \pi ^2}{4096}\nonumber \\
   &&-\frac{25 \pi ^4}{1024}-\frac{167 \pi ^6}{30240}-\frac{79 \pi ^8}{453600}+\frac{79 \pi
   ^{10}}{498960}) S^5+(-\frac{477}{1024}+\frac{271 \pi ^2}{6144}-\frac{11 \pi
   ^4}{960}+\frac{53 \pi ^6}{8640}\nonumber \\
   &&+\frac{16 \pi ^8}{2835}-\frac{67 \pi ^{10}}{106920})
   S^4+(\frac{195}{1024}-\frac{679 \pi ^2}{1536}+\frac{139 \pi ^4}{1440}+\frac{311 \pi
   ^6}{5040}-\frac{236 \pi ^8}{14175}+\frac{197 \pi ^{10}}{187110})
   S^3\nonumber \\
   &&+(\frac{21}{128}-\frac{139 \pi ^2}{192}+\frac{151 \pi ^4}{720}-\frac{32 \pi ^6}{315}+\frac{22
   \pi ^8}{1575}-\frac{4 \pi ^{10}}{6237}) S^2+\frac{S}{2},\\
P_9&=& -\frac{42639
   S^8}{262144}+(\frac{24345}{32768}-\frac{713 \pi ^2}{4096}-\frac{659 \pi ^4}{61440}-\frac{31 \pi
   ^6}{48384}+\frac{17 \pi ^8}{345600}+\frac{\pi ^{10}}{47520}+\frac{691 \pi ^{12}}{309582000})
   S^7\nonumber \\
   &&+(-\frac{196863}{131072}+\frac{11205 \pi ^2}{16384}+\frac{8957 \pi ^4}{368640}+\frac{7831 \pi
   ^6}{967680}+\frac{91 \pi ^8}{64800}-\frac{712373 \pi ^{12}}{20432412000})
   S^6\nonumber \\
   &&+(\frac{99543}{65536}-\frac{5045 \pi ^2}{8192}+\frac{227 \pi ^4}{4096}+\frac{4537 \pi
   ^6}{483840}-\frac{7813 \pi ^8}{1814400}-\frac{227 \pi ^{10}}{138600}+\frac{76627 \pi
   ^{12}}{378378000}) S^5\nonumber \\
   &&+(-\frac{12969}{16384}+\frac{623 \pi ^2}{1024}-\frac{1639 \pi
   ^4}{23040}-\frac{155 \pi ^6}{4032}-\frac{43 \pi ^8}{1680}+\frac{7913 \pi ^{10}}{935550}-\frac{102253
   \pi ^{12}}{182432250}) S^4\nonumber \\
   &&+(\frac{51}{1024}+\frac{689 \pi ^2}{1536}-\frac{863 \pi
   ^4}{3840}-\frac{6397 \pi ^6}{30240}+\frac{11867 \pi ^8}{113400}-\frac{1058 \pi
   ^{10}}{66825}+\frac{159403 \pi ^{12}}{212837625}) S^3\nonumber \\
   &&+(-\frac{43}{256}+\frac{57 \pi
   ^2}{64}-\frac{181 \pi ^4}{480}+\frac{503 \pi ^6}{1260}-\frac{152 \pi ^8}{1575}+\frac{64 \pi
   ^{10}}{6237}-\frac{5528 \pi ^{12}}{14189175}) S^2-\frac{S}{2},\\
P_{10}&=& \frac{716283
   S^9}{4194304}+(-\frac{2091669}{2097152}+\frac{67109 \pi ^2}{262144}+\frac{899 \pi
   ^4}{49152}+\frac{5447 \pi ^6}{3096576}+\frac{277 \pi ^8}{2764800}-\frac{\pi ^{10}}{84480}\nonumber \\
   &&-\frac{691
   \pi ^{12}}{176904000}-\frac{\pi ^{14}}{2721600}) S^8+(\frac{692223}{262144}-\frac{19251 \pi
   ^2}{16384}-\frac{8141 \pi ^4}{491520}-\frac{10967 \pi ^6}{1290240}-\frac{69043 \pi
   ^8}{29030400}\nonumber \\
   &&-\frac{26197 \pi ^{10}}{79833600}+\frac{712373 \pi ^{12}}{81729648000}+\frac{10981 \pi
   ^{14}}{1459458000}) S^7+(-\frac{1006731}{262144}+\frac{718259 \pi ^2}{393216}-\frac{5227
   \pi ^4}{36864}\nonumber \\
   &&-\frac{29413 \pi ^6}{1105920}-\frac{25741 \pi ^8}{14515200}+\frac{3727 \pi
   ^{10}}{1871100}+\frac{32947 \pi ^{12}}{77837760}-\frac{521273 \pi ^{14}}{8756748000})
   S^6+(\frac{203763}{65536}-\frac{111121 \pi ^2}{65536}\nonumber 	\\
   &&+\frac{5429 \pi ^4}{40960}-\frac{179 \pi
   ^6}{1290240}+\frac{48613 \pi ^8}{1451520}+\frac{7013 \pi ^{10}}{950400}-\frac{17579927 \pi
   ^{12}}{5108103000}+\frac{349303 \pi ^{14}}{1459458000})
   S^5\nonumber \\
   &&+(-\frac{18357}{16384}+\frac{3779 \pi ^2}{16384}+\frac{2201 \pi ^4}{18432}+\frac{145009 \pi
   ^6}{967680}+\frac{165811 \pi ^8}{1814400}-\frac{200791 \pi ^{10}}{2993760}\nonumber \\
   &&+\frac{1027603 \pi
   ^{12}}{94594500}-\frac{7441 \pi ^{14}}{14215500}) S^4+(\frac{741}{4096}-\frac{21 \pi
   ^2}{32}+\frac{1633 \pi ^4}{3840}+\frac{4081 \pi ^6}{5760}-\frac{8137 \pi ^8}{15120}\nonumber \\
   &&+\frac{45013 \pi
   ^{10}}{311850}-\frac{60548 \pi ^{12}}{3869775}+\frac{54298 \pi ^{14}}{91216125})
   S^3+(\frac{85}{512}-\frac{811 \pi ^2}{768}+\frac{307 \pi ^4}{480}-\frac{817 \pi
   ^6}{560}+\frac{284 \pi ^8}{525}\nonumber \\
   &&-\frac{208 \pi ^{10}}{2079}+\frac{11056 \pi ^{12}}{1289925}-\frac{16
   \pi ^{14}}{57915}) S^2+\frac{S}{2},\\
P_{11}&=& -\frac{1514943
   S^{10}}{8388608}+(\frac{21799467}{16777216}-\frac{186593 \pi ^2}{524288}-\frac{35769 \pi
   ^4}{1310720}-\frac{9701 \pi ^6}{3096576}-\frac{7099 \pi ^8}{22118400}-\frac{169 \pi
   ^{10}}{10644480}\nonumber \\
   &&+\frac{691 \pi ^{12}}{254016000}+\frac{\pi ^{14}}{1360800}+\frac{3617 \pi
   ^{16}}{58296672000}) S^9+(-\frac{36451737}{8388608}+\frac{255955 \pi ^2}{131072}-\frac{1193
   \pi ^4}{393216}+\frac{6433 \pi ^6}{1105920}\nonumber \\
   &&+\frac{623593 \pi ^8}{232243200}+\frac{5729 \pi
   ^{10}}{8870400}+\frac{5891749 \pi ^{12}}{81729648000}-\frac{10981 \pi
   ^{14}}{2918916000}-\frac{11128631 \pi ^{16}}{6947020080000})
   S^8\nonumber \\
   &&+(\frac{8850417}{1048576}-\frac{139205 \pi ^2}{32768}+\frac{20249 \pi ^4}{61440}+\frac{91559
   \pi ^6}{1935360}+\frac{75661 \pi ^8}{7257600}-\frac{13829 \pi ^{10}}{79833600}-\frac{20085073 \pi
   ^{12}}{27243216000}\nonumber \\
   &&-\frac{190951 \pi ^{14}}{1915538625}+\frac{3888667 \pi ^{16}}{236830230000})
   S^7+(-\frac{5148435}{524288}+\frac{2033689 \pi ^2}{393216}-\frac{16397 \pi
   ^4}{36864}+\frac{289517 \pi ^6}{7741440}\nonumber \\
   &&-\frac{92989 \pi ^8}{4838400}-\frac{59173 \pi
   ^{10}}{3326400}-\frac{1495889 \pi ^{12}}{1135134000}+\frac{74191597 \pi
   ^{14}}{61297236000}-\frac{348304631 \pi ^{16}}{3907698795000})
   S^6+(\frac{3259851}{524288}\nonumber \\
   &&-\frac{192831 \pi ^2}{65536}+\frac{2503 \pi ^4}{15360}-\frac{7007 \pi
   ^6}{184320}-\frac{2386849 \pi ^8}{14515200}-\frac{269257 \pi ^{10}}{19958400}+\frac{27220307 \pi
   ^{12}}{851350500}\nonumber \\
   &&-\frac{13439483 \pi ^{14}}{2357586000}+\frac{21550687 \pi ^{16}}{76621545000})
   S^5+(-\frac{219645}{131072}+\frac{19109 \pi ^2}{16384}-\frac{7933 \pi ^4}{23040}-\frac{173951 \pi
   ^6}{322560}-\frac{983261 \pi ^8}{3628800}\nonumber \\
   &&+\frac{2085959 \pi ^{10}}{4989600}-\frac{15685156 \pi
   ^{12}}{127702575}+\frac{104438203 \pi ^{14}}{7662154500}-\frac{7824683 \pi ^{16}}{15029610750})
   S^4+(-\frac{627}{16384}\nonumber \\
   &&+\frac{545 \pi ^2}{768}-\frac{2051 \pi ^4}{2880}-\frac{93461 \pi
   ^6}{40320}+\frac{372437 \pi ^8}{151200}-\frac{650801 \pi ^{10}}{623700}+\frac{8246092 \pi
   ^{12}}{42567525}-\frac{10522118 \pi ^{14}}{638512875}\nonumber\\
   &&+\frac{17121956 \pi ^{16}}{32564156625})
   S^3+(-\frac{171}{1024}+\frac{313 \pi ^2}{256}-\frac{47 \pi ^4}{45}+\frac{2861 \pi
   ^6}{560}-\frac{1879 \pi ^8}{700}+\frac{536 \pi ^{10}}{693}\nonumber \\
   &&-\frac{176896 \pi ^{12}}{1576575}+\frac{464
   \pi ^{14}}{57915}-\frac{115744 \pi ^{16}}{516891375}) S^2-\frac{S}{2}.
\ea

\section{Leading order coefficients $c_{n,k}$}
\label{app:LOc}

\ba
c_{1,0} &=&  -2 S (S+2), \\
c_{1,1} &=&  8 S (5 S-6), \\
c_{2,0} &=&  \frac{1}{2} S
   \left(3 S^2+2 S+8\right), \\
   c_{2,1} &=&  -10 S \left(5 S^2+10
   S-32\right), \\
   c_{2,2} &=&  8 S \left(73 S^2-242 S+200\right), \\
   c_{2,3} &=&  32 S
   \left(13 S^2-30 S+16\right), \\
   c_{3,0} &=&  -\frac{1}{16} S \left(21 S^3-6 S^2+24
   S+64\right), \\
   c_{3,1} &=&  \frac{1}{4} S \left(237 S^3+330 S^2+1304
   S-5312\right), \\
   c_{3,2} &=&  -14 (S-2) S \left(83 S^2+276
   S-1008\right), \\
   c_{3,3} &=&  8 S \left(1169 S^3-8182 S^2+17720
   S-12320\right), \\
   c_{3,4} &=&  16 S \left(1167 S^3-4962 S^2+6984
   S-3200\right), \\
   c_{3,5} &=&  64 (S-2) S^2 (S+4), \\c_{4,0} &=&  \frac{1}{128} S
   \left(159 S^4-198 S^3+192 S^2+160 S+512\right), \\
   c_{4,1} &=&  -\frac{1}{32} S
   \left(2241 S^4+1118 S^3+12272 S^2+30368 S-149504\right), \\
   c_{4,2} &=& 
   \frac{3}{8} S \left(4905 S^4+8294 S^3+9056 S^2-456352
   S+730112\right), \\
   c_{4,3} &=&  -\frac{3}{2} S \left(17479 S^4-5662 S^3-661520
   S^2+2242144 S-2078720\right), \\
   c_{4,4} &=&  2 S \left(72669 S^4-1004274
   S^3+4214400 S^2-7226912 S+4452864\right), \\
   c_{4,5} &=&  8 S \left(78045
   S^4-539514 S^3+1412464 S^2-1650144 S+720896\right), \\
   c_{4,6} &=&  32 S
   \left(3785 S^4-20474 S^3+41120 S^2-36512 S+12288\right), \\
   c_{4,7} &=&  -128
   (S-2) S^2 (S+4) (7 S-12).
\ea

\section{Derivation of the $\mc O(P^{2})$ expansion of the Bethe Ansatz}
\label{app:NLO}

The first problem in treating (\ref{eq:master}) is the evaluation of  the following quantity (sum over both $i$ and $j$)
\be
\sum_{j\neq i}\frac{1}{x-x_{i}}\frac{1}{(x_{i}-x_{j})^{3}}.
\ee
It is convenient to define 
\be
F(x, a) = \sum_{j\neq i}\frac{1}{x-x_{i}}\frac{1}{x_{i}-x_{j}+a}.
\ee
Exploiting partial fraction decomposition, we find
\be
F(x,a) = \sum_{j\neq i}\left[
\frac{1}{x-x_{i}}\frac{1}{x+a-x_{j}}+\frac{1}{x+a-x_{j}}\frac{1}{a+x_{i}-x_{j}}
\right],
\ee
and, after some manipulation, 
\be
F(x,a) = G(x)\,G(x+a)+\frac{G(x+a)-G(x)}{a}-F(x+a,-a).
\ee
Expanding in $a$ at order $a^{3}$, we find the identity
\be
\sum_{j\neq i}\frac{1}{x-x_{i}}\frac{1}{(x_{i}-x_{j})^{3}}=-\frac{1}{2}\frac{d}{dx}\sum_{j\neq i}\frac{1}{x-x_{i}}\frac{1}{(x_{i}-x_{j})^{2}}-\frac{1}{24}\,G'''(x)-\frac{1}{4}\,G'(x)^{2}.
\ee
So the problem is reduced to the calculation of 
\be
H(x)=\sum_{j\neq i}\frac{1}{x-x_{i}}\frac{1}{(x_{i}-x_{j})^{2}}.
\ee
This quantity can be extracted by taking (\ref{eq:master}), multiplying by $\frac{1}{x-x_{i}}\frac{1}{x_{i}-x_{m}}$
and summing over $i,m$ with $i\neq m$. Using $G(x)=-G(-x)$ we have (looking at the poles)
\be
\sum_{j\neq i}\frac{x_{i}}{x_{i}^{2}+\varepsilon^{2}}\frac{1}{x-x_{i}}\frac{1}{x_{i}-x_{m}} = 
\frac{1}{2}\,\frac{x}{x^{2}+\varepsilon^{2}}\,\bigg(
G^{2}(x)+G'(x)-G^{2}(i\,\varepsilon)-G'(i\,\varepsilon)
\bigg).
\ee
Taking $\varepsilon=0$, we also find (using $G(0)=0$)
\be
\sum_{j\neq i}\frac{1}{x_{i}}\frac{1}{x-x_{i}}\frac{1}{x_{i}-x_{m}} = 
 \frac{1}{2\,x}\,\bigg(
G^{2}(x)+G'(x)-G'(0).
\bigg) 
\ee
Finally, 
\ba
\mathop{\mathop{\sum_{i}}_{m\neq i}}_{j\neq i}\frac{1}{x-x_{i}}\frac{1}{x_{i}-x_{m}}\frac{1}{x_{i}-x_{j}} &=& 
\mathop{\sum_{i}}_{j\neq i}\frac{1}{x-x_{i}}\frac{1}{(x_{i}-x_{j})^{2}} +
\mathop{\sum_{ijm}}_{all\ distinct}\frac{1}{x-x_{i}}\frac{1}{x_{i}-x_{m}}\frac{1}{x_{i}-x_{j}}\nonumber \\
&=& H(x)+\frac{1}{3}G(x)^{3}+G(x)G'(x)+\frac{1}{3}G''(x).
\ea
Putting together all the pieces, and defining $J^{3}H(J x)=\mc H(x)$, we obtain (\ref{eq:master-NLO}).

\section{Next-to-leading order coefficients $c_{n,k; 1}$}
\label{app:NLOc}

\ba
c_{1,0;1} &=&  0,\\
c_{1,1;1} &=&  0,\\
c_{2,0;1} &=&  -\frac{2}{3} (S-6) S,\\
c_{2,1;1} &=&  -\frac{8}{3} S
   (S+10),\\
   c_{2,2;1} &=&  \frac{32}{3} S (5 S+2),\\
   c_{2,3;1} &=&  -\frac{128}{3} S (3 S-2),\\
   c_{3,0;1} &=&  \frac{1}{3}
   S \left(S^2+18 S-72\right),\\
   c_{3,1;1} &=&  -\frac{4}{3} S \left(13 S^2-138 S+96\right),\\
   c_{3,2;1} &=&  -32 S
   \left(5 S^2+34 S-56\right),\\
   c_{3,3;1} &=&  \frac{128}{3} S \left(47 S^2-38 S-48\right),\\
   c_{3,4;1} &=& 
   -\frac{256}{3} S (3 S-4) (17 S-18),\\
   c_{3,5;1} &=&  -\frac{1024}{3} (S-2) S^2,\\
   c_{4,0;1} &=&  \frac{1}{48} S
   \left(S^3-326 S^2-1112 S+4800\right),\\
   c_{4,1;1} &=&  \frac{1}{12} S \left(205 S^3+2978 S^2-33048
   S+46400\right),\\
   c_{4,2;1} &=&  -S \left(329 S^3-8454 S^2+11560 S+4224\right),\\
   c_{4,3;1} &=&  -\frac{4}{3} S
   \left(5151 S^3+25926 S^2-119752 S+104832\right),\\
   c_{4,4;1} &=&  \frac{16}{3} S \left(11619 S^3-31666
   S^2-648 S+38848\right),\\
   c_{4,5;1} &=&  -64 (S-2) S \left(1619 S^2-5660 S+4448\right),\\
   c_{4,6;1} &=& 
   -\frac{256}{3} S \left(841 S^3-2918 S^2+2984 S-768\right),\\
   c_{4,7;1} &=&  \frac{1024}{3} (S-2) S^2 (11
   S-12).
   \ea

\section{Convergence of the small winding expansion}
\label{app:convergence}

The convergence of the integral representations discussed in Sec.~(\ref{sec:match})
can be proved by the methods that we illustrate in a prototypical example. 
Let us consider the quantity
\be
F(z)=\int_0^1 \frac{1-zx/\pi}{\cos z x/2} dx 
\ee
and its value at $z=\pi$
\be
F(\pi) = \frac{8\,C}{\pi},\qquad C = \mbox{Catalan's constant} = 0.915965594\dots.
\ee

Consider $G(w) = \frac{1-w/\pi}{\cos w/2}$. This function is holomorphic
for $-\pi < \mbox{Re}(w) < 3\pi$. We can subtract the pole in $-\pi$ and consider 
\be
H(w)=G(w)-\frac{4}{w+\pi}.
\ee
This function is holomorphic for $-3\pi < \mbox{Re}(w) < 3\pi$. Hence,
\be
H(w)=\sum_{n=0}^{\infty} a_n\,w^n,\qquad |x| \le 2\pi.
\ee
This implies
\be
\int_{0}^{1}H(zx) dx = \sum_{n=0}^{\infty} \frac{a_n}{n+1}\,z^n,\qquad |z| \le 2\pi.
\ee
Finally
\be
F(z) = \int_0^1 H(zx) dx +4/z \log (1+z/\pi), 
\ee
and we have  $F(\pi)$ as a convergence series in $\pi$
\ba
F(\pi) &=& \sum_{n=0}^{\infty} \frac{a_n}{n+1} \pi^n +\frac{4}{\pi}\, \log 2 \nonumber \\
&=&
\frac{1}{2}+\frac{\pi^2}{96}+\frac{\pi^4}{2304}+\frac{61 \pi^6}{2580480}+\frac{277
   \pi^8}{185794560}+\frac{50521 \pi^{10}}{490497638400}+\cdots
\ea

\bibliography{AC-Biblio}{}
\bibliographystyle{JHEP}

\end{document}